\documentclass[amssymb,aps,floatfix,floats,preprintnumbers,prl,twocolumn,superscriptaddress,groupedaddress]{revtex4}
\usepackage{rcs}
\usepackage{color,graphicx}
\usepackage{dcolumn}
\usepackage{bm}
\usepackage[normalem]{ulem}
\usepackage{chngcntr}
\usepackage[thinspace,thinqspace]{SIunits}
\usepackage{amsmath}
\usepackage{natbib}
\usepackage{doi}
\usepackage{hyperref}

\newcommand{\et}{\textit{et al.}}
\newcommand{\muB}{$\mu_{\textrm{B}}$}
\newcommand{\kB}{$k_{\textrm{B}}$}
\newcommand{\mueff}{$\mu_{\textrm{eff}}$}

\newcommand{\YRS}{YbRh$_{2}$Si$_{2}$}
\newcommand{\YPS}{YbPt$_{2}$Sn}
\newcommand{\YPI}{YbPt$_{2}$In}

\newcommand{\YCA}{YbCu$_{4.6}$Au$_{0.4}$}
\newcommand{\YCAx}{YbCu$_{5-x}$Au$_{x}$}
\newcommand{\YCN}{YbCu$_4$Ni}
\newcommand{\TK}{$T_{\textrm{K}}$}
\newcommand{\TRKKY}{$T_{\textrm{RKKY}}$}
\newcommand{\Tm}{$T_{\textrm{m}}$}
\begin{document}
\preprint{APS/123-QED}
\title{Electronuclear Quantum Criticality}
\author{J. Banda}
\author{D. Hafner}
\affiliation{Max Planck Institute for Chemical Physics of Solids, D-01187 Dresden, Germany}
\author{J. F. Landaeta}
\affiliation{Max Planck Institute for Chemical Physics of Solids, D-01187 Dresden, Germany}
\affiliation{Institute of Solid State and Materials Physics, TU Dresden, 01069, Dresden, Germany}
\author{E. Hassinger}
\affiliation{Institute of Solid State and Materials Physics, TU Dresden, 01069, Dresden, Germany}
\affiliation{Max Planck Institute for Chemical Physics of Solids, D-01187 Dresden, Germany}
\author{K. Mitsumoto}
\affiliation{Liberal Arts and Sciences, Toyama Prefectural University, Imizu, Toyama 939-0398, Japan}
\author{M. Giovannini}
\affiliation{Department of Chemistry and Industrial Chemistry (DCCI), University of Genova, 16100 Genova, Italy}
\author{J. G. Sereni}
\affiliation{Department of Physics, CAB-CNEA, CONICET, 8400 San Carlos de Bariloche, Argentina}
\author{C. Geibel}
\author{M. Brando}
\email[Corresponing author:~]{brando@cpfs.mpg.de}
\affiliation{Max Planck Institute for Chemical Physics of Solids, D-01187 Dresden, Germany}
\date{\today}
\begin{abstract}
We present here a rare example of electronuclear quantum criticality in a metal. The compound \YCA\ is located at an unconventional quantum critical point (QCP). In this material the relevant Kondo and RKKY exchange interactions are very weak, of the order of 1\,K. Furthermore, there is strong competition between antiferromagnetic and ferromagnetic correlations, possibly due to geometrical frustration within the fcc Yb sublattice. This causes strong spin fluctuations which prevent the system to order magnetically. Because of the very low Kondo temperature the Yb$^{3+}$ $4f$-electrons couple weakly with the conduction electrons allowing the coupling to the nuclear moments of the $^{171}$Yb and $^{173}$Yb isotopes to become important. Thus, the quantum critical fluctuations observed at the QCP do not originate from purely electronic states but from entangled electronuclear states. This is evidenced by the anomalous temperature and field dependence of the specific heat at low temperatures.
\end{abstract}
\maketitle
Since its birth concept in 1976~\cite{hertz1976} quantum criticality (QC) has become an important topic in modern physics. QC has been investigated in a large variety of systems and materials (see Refs.~\cite{mathur1998,sachdev2000,vojta2003,senthil2004,coleman2005,sachdev2008,broun2008,gegenwart2008,cejnar2010,shibauchi2014,rowley2014,brando2016,vojta2018,broholm2020} and references therein) in which the nature of quantum fluctuations was studied. Such fluctuations are electronic in nature and their effects measurable at temperature well above 10\,mK~\cite{custers2003}, even though the QCP is technically located at $T = 0$. The effect of nuclear degrees of freedom is commonly neglected. This is because the dipole hyperfine coupling energy $H_{hf} = A \mathbf{I} \cdot \mathbf{J}$ of nuclear spin $I$ and total electron angular momentum $J$ is very low, of the order of 1\,mK or lower. However, in lanthanide - in which the $4f$-electrons are located close to the nucleus - $A$  can be sufficiently large to influence the magnetic ground states~\cite{kondo1961}. Prime examples are Pr$^{3+}$- or Ho$^{3+}$-based systems~\cite{steinke2013,bitko1996,libersky2021,wendl2022} in which $A$ has the exceptionally high values of 52 and 39\,mK, respectively~\cite{kondo1961,bleaney1963}.

In Yb$^{3+}$-based systems nuclear moments may also participate to the ground state because of the large $A \approx 40$\,mK~\cite{flouquet1978}, see Supplementar Material (SM). However, in metals it strongly depends on the Kondo effect. Usually the Kondo temperature is much larger, \TK\ $\gg A$. Then the $4f$-electrons strongly couple to the conduction electrons by the Kondo exchange interaction and not to the nuclear moments; consequently the nuclear moments are subjected to an hyperfine field which is proportional to the $4f$-electron magnetization. A typical example is \YRS\ with \TK\ $\approx 25$\,K~\cite{knebel2006,knapp2023}. Only when \TK\ $\ll A$ the nuclear and $4f$-moments couple first to form a mixed electronuclear ground state; the Kondo exchange is a perturbation~\cite{flouquet1978}. One example is the metal YbFe$_{2}$Si$_{2}$~\cite{noakes1983}, but this effect was well observed in paramagnetic salts~\cite{nowik1968}.

These two regimes are known as 'fast relaxation' (strong coupling) and 'slow relaxation' (weak coupling), respectively. This is because the decoupling of $I$ and $J$ is attained through rapid fluctuations ($\tau_{J}$) of $J$ with respect to the nuclear Larmor period ($\tau_{n} \approx h/A$\kB). Hence, all depends on $\tau_{J} $ being larger or smaller than $\tau_{n}$. This can be precisely measured by M\"ossbauer spectroscopy~\cite{nowik1968,bonville1986}:  relaxations rates are typically between $\tau_{J} \approx 10^{-8}$\,s (slow relaxation) to $\tau_{J} \approx 10^{-12}$\,s (fast relaxation)~\cite{nowik1968}. Relaxation rates in Kondo systems can be estimated by $\tau = h/$\kB\TK, whereas in magnetically ordered systems, the RKKY interaction leads to a spin-spin relaxation rate $\tau = h/$\kB$T_{m}$ where \Tm\ is the ordering temperature.

The reason why in QC $4f$-electron metallic systems the role of nuclear moments have been always neglected is because, following the Doniach principle~\cite{doniach1977}, either a strong Kondo effect prevents them to couple to the $4f$-moments or, if \TRKKY\ $>$ \TK\ the system orders magnetically. In general, at the QCP we have the situation with a large \TK\ $>$ \TRKKY\ and a fast relaxation ($\tau_{J} \ll \tau_{n}$).

Here, we present a rare example of electronuclear quantum criticality in a metal. We show that the compound \YCA\ is located at an unconventional QCP which results from very weak Kondo and RKKY exchange interactions as well as from a strong competition between antiferromagnetic (AFM) and ferromagnetic (FM) correlations. Thus, despite a low Kondo temperature of about 2\,K, this system does not magnetically order, instead strong spin fluctuations are observed down to 20\,mK. The Yb$^{3+}$ $4f$-electrons couple weakly with the conduction electrons, but strongly with the nuclear spins of the $^{171}$Yb and $^{173}$Yb isotopes because of the slow relaxation driven by spin fluctuations at the QCP. Thus, these QC fluctuations do not originate from purely electronic states but from electronic and nuclear states coupled by the hyperfine interaction. This is evidenced by the temperature and field dependencies of the specific heat $C(T,B)$.     

\begin{figure}[t]
	\begin{center}
		\includegraphics[width=\columnwidth]{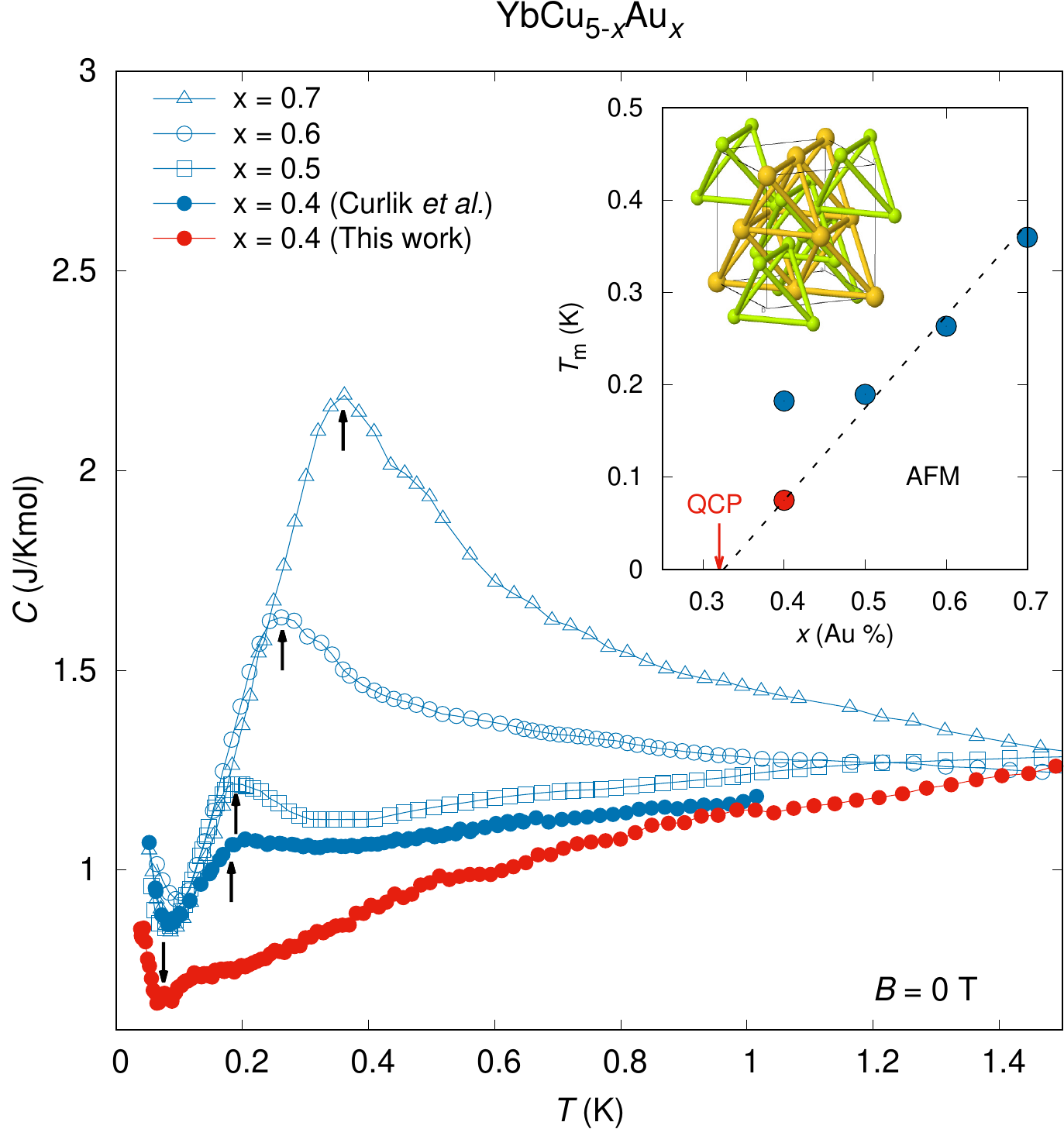}
		\caption{$T$-dependence of the specific heat of five samples of \YCAx\ with $0.4 \leq x \leq 0.7$. The blue points are taken from C\v{u}rl\'{i}k \textit{et al.}~Ref.~\cite{curlik2014}, while the red points are from a new sample measured by us. The arrows mark the temperature \Tm\ of the maxima below which AFM ordering was found. Inset: \Tm\ plotted over the nominal Au content $x$. The picture illustrates the cubic AuBe$_{5}$-type structure with emphasis on the tetrahedra formed by the Yb atoms (yellow).}
		\label{fig1}
	\end{center}
\end{figure}

\YCAx\ crystallizes in the cubic C15b AuBe$_{5}$-type structure (space group F-43m) with Yb atoms located at the vertices of edge-sharing tetrahedra~\cite{yoshimura2001,giovannini2005} with cubic local symmetry (see inset of Fig.~\ref{fig1}). Au occupies the Wyckoff site 4c with cubic symmetry, but Cu is on the 16e site with trigonal simmetry. In \YCA\ the additional 0.6 of Cu goes into the 4c position~\cite{giovannini2005}.

The Yb atoms are in the trivalent state Yb$^{3+}$. The cubic cristalline electric field (CEF) splits the $J = 7/2$ energy levels into a $\Gamma_{8}$ quartet and two Kramers doublets, $\Gamma_{6}$ and $\Gamma_{7}$. Although a $\Gamma_7$ was suggested to be the ground state~\cite{severing1990,bonville1992}, the value of the magnetization at 7\,T of about 1.5\,\muB~\cite{sereni2018} (cf. inset of Fig.~\ref{fig4}) suggests a possible $\Gamma_{6}$, since $M_{sat}(\Gamma_{6}) = 1.33$\,\muB\ and $M_{sat}(\Gamma_{7}) = 1.72$\,\muB. The alloy \YCAx\ has been reported to show AFM ordering below 1\,K and a continuous evolution from the ordered state to a dense Kondo state and a QCP with non-Fermi liquid properties at $x = 0.2 - 0.4$~\cite{yoshimura2001,galli2002}. 
Near the QCP a strong enhancement of the electronic specific heat coefficient $C(T)/T \propto T^{-n}$ with $n > 1$ was reported~\cite{curlik2014} resulting in a large entropy accumulation below 4\,K, similar to what was observed in other Yb-based cubic systems like \YPI~\cite{gruner2014} or \YCN~\cite{sereni2018}.

The low-$T$ properties of compounds with $0.4 \leq x \leq 0.7$ were extensively investigated by C\v{u}rl\'{i}k \textit{et al.} down to 0.05\,K~\cite{curlik2014}. They found a continuous suppression of the long-range AFM ordering with decreasing $x$, with a N\'{e}el temperature $T_{N} \approx 0.37$\,K for nominal $x = 0.7$ and $T_{N} \approx 0.18$\,K with $x = 0.4$. This is shown in Fig.~\ref{fig1}. Thus, the QCP is located just near the Au content $x = 0.4$ as previously suggested by Yoshimura \textit{et al.}~\cite{yoshimura2001}. The sample near the QCP showed non-Fermi-liquid behavior, i.e. a strong increase of the resistivity and $C(T)/T \propto T^{-0.89}$ down to 0.2\,K below which AFM ordering was observed. C\v{u}rl\'{i}k \textit{et al.} have estimated the Kondo temperature for this system and found \TK\ $< 2$\,K. So, an ordering temperature of 0.2\,K with such a low Kondo temperature is rare and points to possible frustration effects. For this reason we decided to investigate a sample prepared by the same group with nominal composition $x = 0.4$ or lower, i.e., close to the QCP.

Fig.~\ref{fig1} illustrates the behavior of our sample compared to those measured by C\v{u}rl\'{i}k \et~\cite{curlik2014}. We have plotted $C(T)$ vs $T$ and the magnetic transition temperatures \Tm\ versus $x$ (inset) together with our data (red points). In all samples a kink at \Tm\ in $C(T)$ vs $T$ is observed. We note that in our sample with $x = 0.4$ a kink at \Tm\ can be barely seen; a closer look suggests that there is a broad hump which would indicate a sort of very weak short-range order or spin glass freezing. We will show later that in our sample a clear signature in the ac-susceptibility is found at \Tm\ = 75\,mK. There is a clear and monotonic decrease of \Tm\ with decreasing $x$ indicated by the dashed line in the inset of Fig.~\ref{fig1}, but the sample with $x = 0.4$ studied by C\v{u}rl\'{i}k \et\ is off the line. This suggests that the nominal composition for this specific sample does not correspond to the real composition. On the other hand, the position of \Tm\ in our sample suggests that its nominal composition $x = 0.4$ is very similar to the real one and the sample is closer to the QCP.

The feature that indicates a phase transition in specific heat in our sample is not sharp indicating a tiny change of entropy. This signature is also affected by the strong increase of $C(T)$ below 0.1\,K which has to be related to the nuclear contribution to $C(T)$ which should dominate at very low $T$. This is, however, remarkable if we assume no static magnetic ordering below 0.1\,K. In fact, Yb and Au are located on sites with cubic symmetry, i.e. zero electric gradient field and therefore no quadrupole contribution to $C(T)$, and the quadrupole contribution of Cu is negligible (see SM). We will see later that this strong increase $C/T \propto 1/T^{2}$ has a different origin.

To investigate the presence of a magnetic phase transition we have measured the $T$ and $B$ dependence of the ac-susceptibility $\chi'(T,B)$ down to 35\,mK. These measurements were carried out with a home-made susceptometer at a modulated ac-field of $40\mu$T and frequency $f = 1127$\,Hz.
\begin{figure}[t]
	\begin{center}
		\includegraphics[width=\columnwidth]{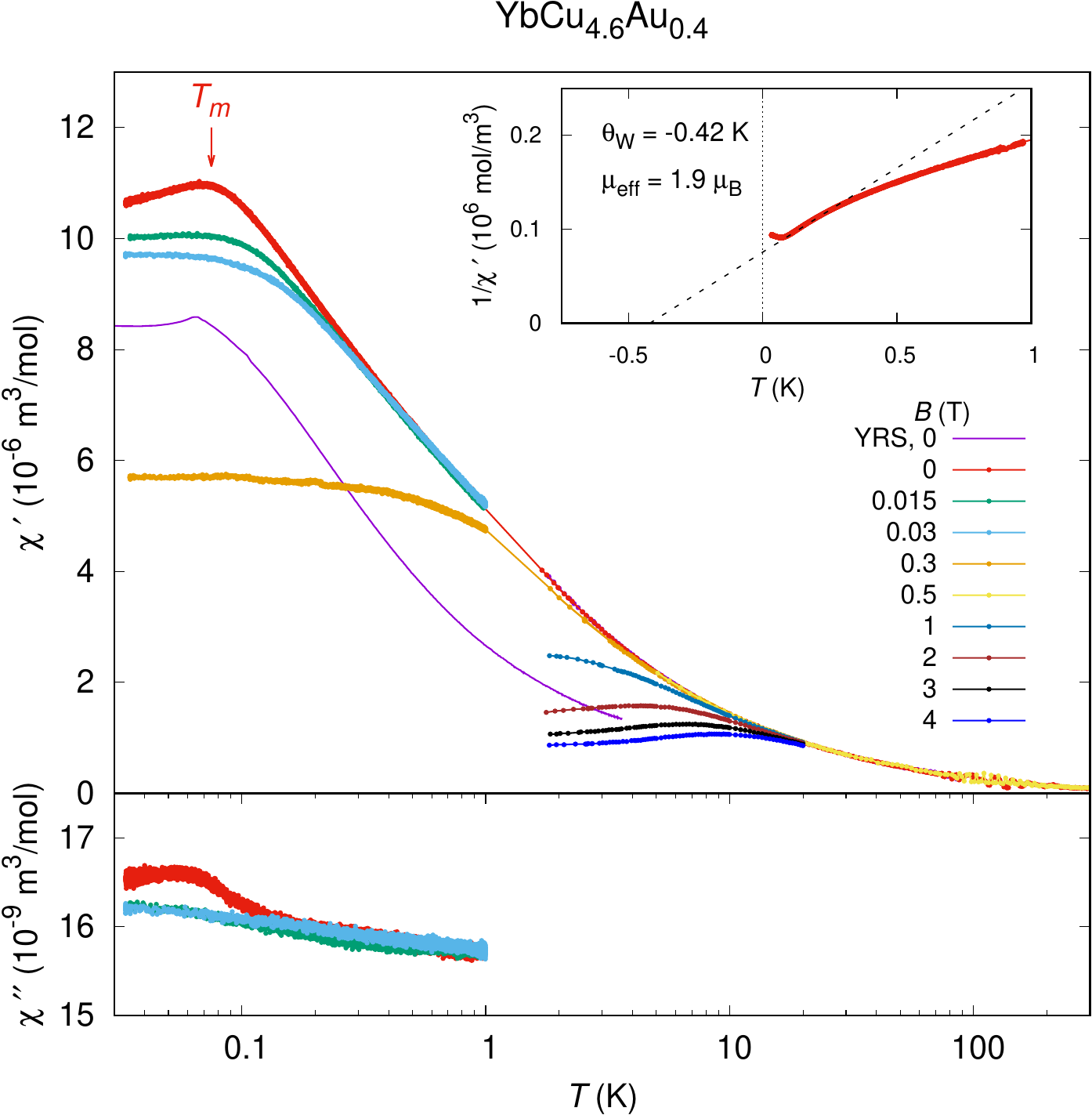}
		\caption{$T$-dependence of the real $\chi'(T)$ and imaginary $\chi''(T)$ part of the ac-susceptibility at different fields. The inset shows the inverse susceptibility data fitted at low $T$ with a Curie-Weiss function. The susceptibility of \YRS\ (YRS) was taken from Ref.~\cite{westerkamp2008}.}
		\label{fig2}
	\end{center}
\end{figure}
Results are summarized in Fig.~\ref{fig2}: In zero field we observe a strong increase of the susceptibility with decreasing $T$, which departs from the CEF Curie-Weiss behavior and reaches at low $T$ high values of about $10^{-5}$\,m$^{3}$/mol. This behavior cannot be ascribed to the CEF susceptibility but is typical of systems very close to a FM instability, as for instance YbRh$_{2}$Si$_{2}$~\cite{hamann2019,brando2016}. To emphasize this point we have plotted $\chi'(T)$ of YbRh$_{2}$Si$_{2}$ on the same graph (purple line). This observation is completely in agreement with a detailed NMR and $\mu$SR study on a $x = 0.6$ sample performed by Carretta \textit{et al.}~\cite{carretta2009}. They have shown that despite AFM ordering strong FM fluctuations are present and are compatible with the self-renormalization (SCR) theory for FM QCP~\cite{carretta2009}. In this study no long-range magnetic ordering was observed but only a slight flattening of the muon spin-lattice relaxation rate was measured below \Tm.

We observe in $\chi'(T)$ a maximum at \Tm\ = 75\,mK which confirms the existence of short-range ordering in our sample. This is accompanied by a small enhancement of $\chi''(T,B)$ indicating dissipative effects. We have analyzed $\chi'(T)^{-1}$ in the inset of Fig.~\ref{fig2}. The Curie-Weis fit on the very low-$T$ data yields a large effective moment \mueff\ = 1.9\,\muB\ which is expected because of the very small Kondo temperature $T_{K} < 2$\,K. This is supported by the $B$-dependence of the magnetization at 2\,K that can be relatively well fitted by a $S = 1/2$ Brillouin function with only the saturation moment of 1.7\,\muB\ as free parameter (see inset of Fig.~\ref{fig4}). The difference between data and fit at low fields can be explained by the presence of intersite correlations already at 2\,K. These correlations are stronger at lower $T$ as it can be seen by the behavior of the magnetization at 45\,mK. Thus, in \YCA\ the Yb moments are only slightly screened by the Kondo effect. 

The small value of the Weiss temperature $\theta_{W} = -0.42$\,K, which indicates dominant AFM exchange interactions, suggests a rather weak coupling between the Yb atoms. An estimation of strength of the exchange interaction can be obtained from Ref.~\cite{curlik2014} by using the measured $4f$-electron entropy $S_{4f}$: taking twice the temperature at which $S_{4f}$ reaches $1/2R\ln 2$ we obtain a value of about 4\,K, i.e. ten times larger than $\theta_{W}$. This suggest that the small value of $\theta_{W}$ is due to the competing AFM and FM interactions.

The kink in $\chi'(T)$ disappears already at very small fields of 15\,mT, which confirms the lack of AFM long-range magnetic ordering. Since this system has a very low Kondo temperature and large effective moment we would expect a much larger critical field. This is confirmed by the field dependence of $\chi'(B)$ measured at 45\,mK between -1.5 and 1.5\,T which does not show any features or hysteresis (see SM). So, the broad transition with a tiny amount of entropy quenched and the small critical field $< 15$\,mT points to weak short-range ordering, the long-range order is prevented by the strong fluctuations. Similar behavior was observed in a closely related system \YCN~\cite{sereni2018}.

To summarize the results so far, \YCA\ is a local moment system with very small Kondo and RKKY exchange interactions. In spite of dominant AFM exchange, strong FM fluctuations are present which results in a weak short-range magnetic ordering below 75\,mK with a tiny amount of quenched entropy. The reason for such behavior is certainly the competition between the AFM and FM interactions which is possibly promoted by the geometrical frustration of the Yb tetrahedra in the cubic crystal structure. Such a peculiar situation results in a large specific heat at low temperature but still can not explain the increase of $C(T)/T$ below 0.1\,K.

\begin{figure}[t]
	\begin{minipage}[t]{\columnwidth}
			\includegraphics[width=\columnwidth]{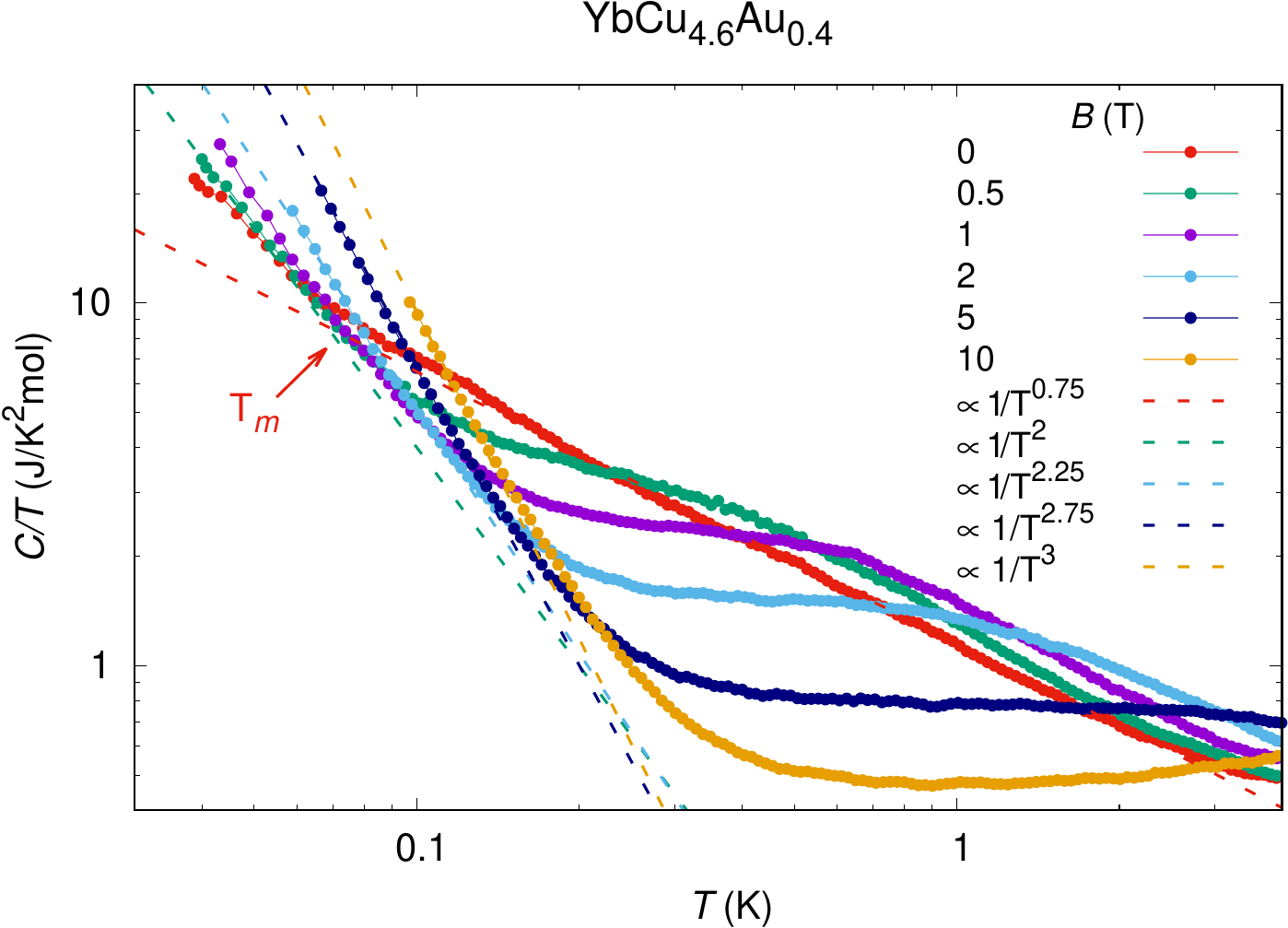}
	\end{minipage}
	\hfill
	\begin{minipage}[t]{\columnwidth}
			\includegraphics[width=\columnwidth]{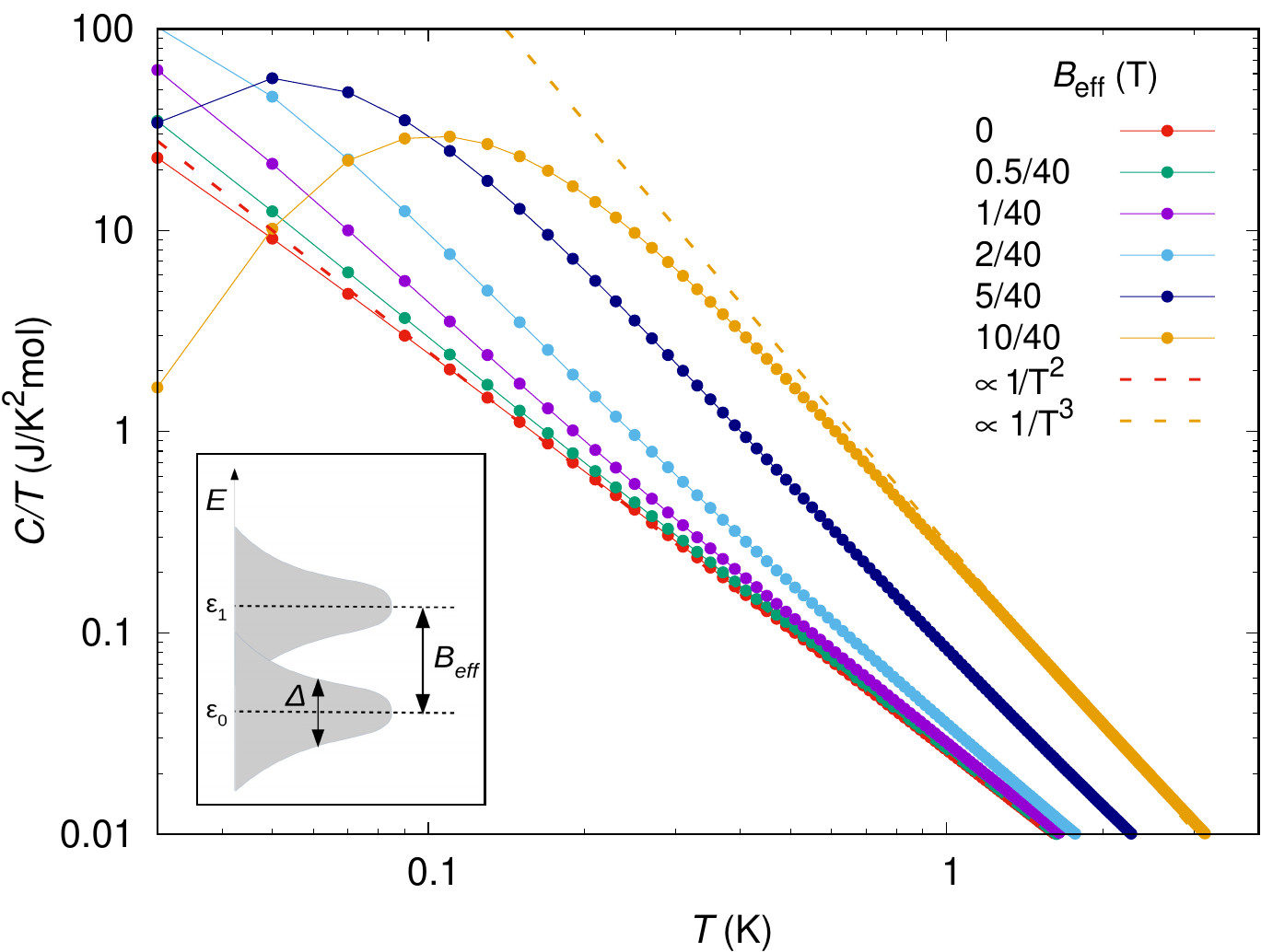}
	\end{minipage}
		\caption{Top: $T$-dependence of the total $C(T)/T$ at selected magnetic fields in a log-log plot. At $B = 0$ $C(T)/T \propto 1/T^{0.75}$ (red dash line) down to 0.2\,K from which it starts to deviate to a higher power-law. The other dashed lines are power-law fits to the lower temperature data to emphasize that the nuclear contribution follows the expected $1/T^{3}$ power-law only at high fields. Bottom: simulation of the behavior observed in $C/T$ vs $T$ with a resonant-level-model (RLM) described in Refs.~\cite{schotte1975,desgranges1982}. Here, we used the following parameters: $S = 1/2$, $g = 2$, $\Delta = 0.01$\,K. We rescaled the magnetic field $B_{eff} = B/40$. The factor 40 is because we used the electron Zeeman energy for the simulation instead of the nuclear one and $\sqrt{ \mu_{B} / \mu_{N} } \approx 42$.}
\label{fig3}
\end{figure}

\begin{figure}[t]
	\begin{center}
		\includegraphics[width=\columnwidth]{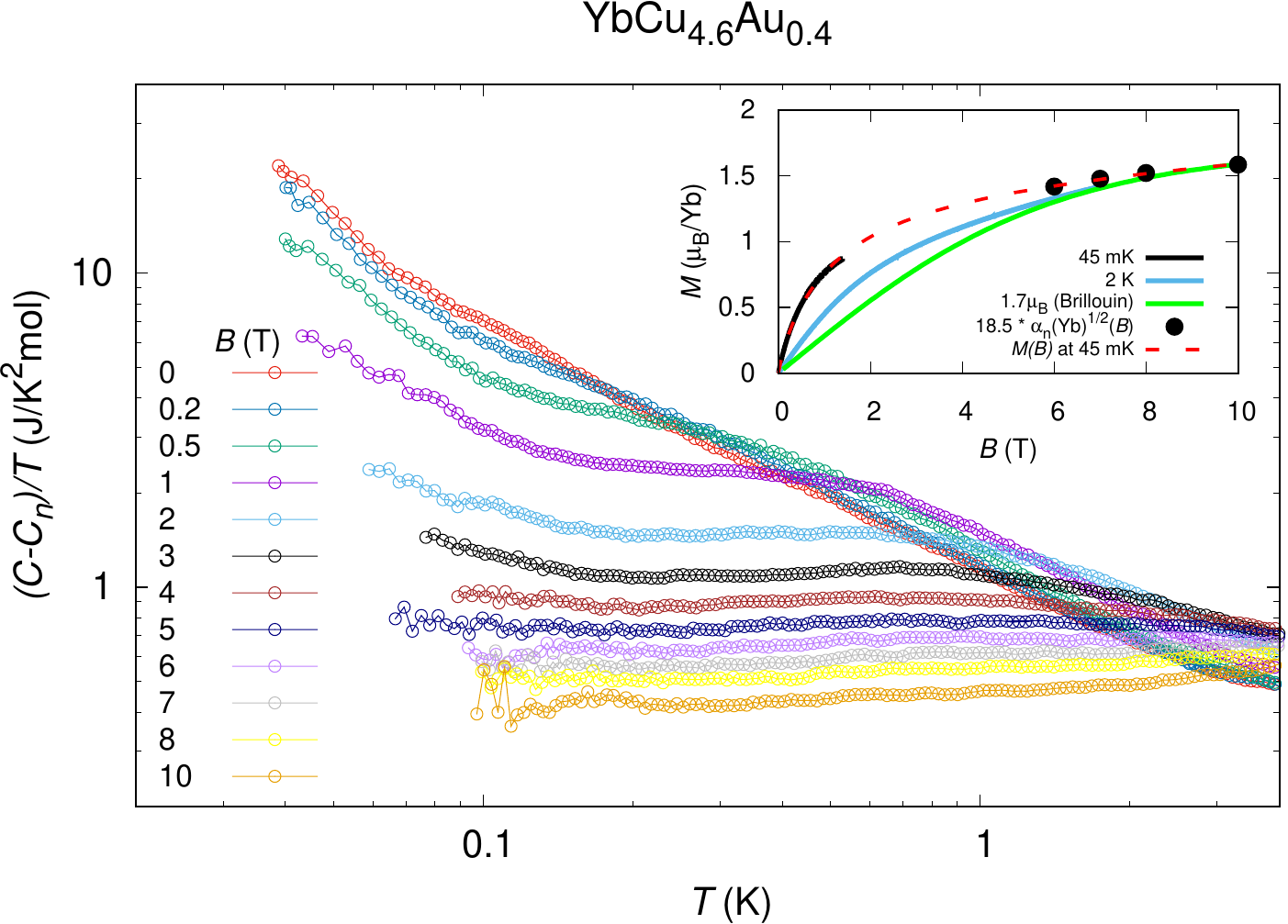}
		\caption{Same plot as in Fig.~\ref{fig3} but with the nuclear specific heat $C_{n}$ subtracted. Here we show all measured fields. Inset: $M(B)$ vs $B$ at 45\,mK and 2\,K. The green line is a $S = 1/2$ Brillouin function calculated with a saturation moment of 1.7\,\muB. The black points are the nuclear contribution coefficients $\alpha_{n}(\mathrm{Yb})$ for the Yb atoms evaluated at 6, 7, 8 and 10\,T. Since $\alpha_{n}(\mathrm{Yb}) \propto M^{2}(B)$, we plotted the square root of the $\alpha_{n}(\mathrm{Yb})$ coefficients multiplied by a constant factor 18.5.}
		\label{fig4}
	\end{center}
\end{figure}

We discuss now the key discovery of our study which is found in the low-$T$ behavior of the specific heat. Fig.~\ref{fig3} shows the specific heat data plotted as $C(T)/T$ vs $T$ in a double logarithmic representation. At $B = 0$, $C(T)/T$ follows a power-law increase $T^{-0.75}$ to lower $T$ similar to that found in Ref.~\cite{curlik2014}. The short-range ordering at \Tm\ can not be identified, possibly because of the very small entropy released at the transition. Below 0.2\,K, $C(T)/T$ starts to increase steeply with a power-law close to $1/T^{2}$. As explained above, this can not be ascribed to just a nuclear quadrupole contribution because of the crystalline cubic structure and the fact that a purely nuclear contribution would follow a Schottky-like behavior, i.e., $C/T \propto 1/T^{3}$. If we had clear static magnetic ordering below \Tm\ then we could explain the $1/T^{2}$ increase by assuming the sum of a decreasing $C/T \propto T$ below \Tm\ due to the ordering and a nuclear contribution $C/T \propto 1/T^{3}$ due to Zeeman splitting of the nuclear levels by the hyperfine field induced by the static $4f$-moments. This has been reported, e.g., for \YPS~\cite{jang2015} or \YCN~\cite{sereni2018}. But, in our case, we do not have static order. This statement is reinforced by the behavior of $C/T$ at the small field $B = 0.2$\,T: The phase transition is suppressed (cf. Fig.~\ref{fig2}) but the behavior of $C(T)/T$ is unchanged (cf. Fig.~\ref{fig4} and Fig.~S1 in SM).
This implies that the magnetic ordering has no effect at all on the behavior of $C(T)/T$. At a higher field, $B = 0.5$\,T, $C(T)/T$ seems to flatten below 0.3\,K but then increases strongly with a similar power law. We observe that the higher the field the clearer is the tendency of the system to enter a Fermi liquid ground state with a constant $C(T)/T$. However, at low temperatures another contribution kicks in, which must be of nuclear origin, and $C(T)/T$ rises steeply again with a power-law exponent that increases with increasing magnetic field.

This is a very peculiar behavior: If only the nuclear contribution was responsible for the large increase of $C(T)/T$ below $T_{m}$ and also at small fields, it should follow a $C/T \propto1/T^{3}$ power law expected for the high-$T$ part of a Schottky peak, as it is normally the case in Yb-based metals in which the nuclear moments are decoupled from the $4f$-moments, e.g., in \YRS~\cite{steppke2010}. Indeed, it does in \YCA, but only at fields larger than 5\,T. We emphasize this behavior in Fig.~\ref{fig3} by plotting data of selected fields with fits to the low-$T$ data indicated by dashed lines. There is a continuous evolution of the power-law exponent $n$ of $C/T \propto 1/T^{n}$ from 2 to 3 with increasing $B$. The data between 1 and 3\,T clearly show a tendency to a constant $C(T)/T$ below 1\,K, followed by a huge increase for $T < 0.3$\,K. Thus, this must be related to the nuclear degrees of freedom.

An elegant way to see whether the enhancement of $C/T$ at low fields can be completely attributed to the nuclear contribution or not is to use the high-field data and the measured magnetization at 45\,mK. We have fitted the low-$T$ data in fields of 6, 7, 8, and 10\,T with $C/T = \gamma + \alpha_{n}/T^{3}$. The total contribution $\alpha_{n}$ consists of $\alpha_{n}(\mathrm{Cu})$ for copper and $\alpha_{n}(\mathrm{Yb})$ for ytterbium. They have different field dependencies, $\alpha_{n}(\mathrm{Cu}) \propto B^{2}$ whereas $\alpha_{n}(\mathrm{Yb}) \propto M^{2}$ because of the very large hyperfine constant $A \approx 110$\,\muB/Yb~\cite{bonville1984,bonville1992,steppke2010} (cf. SM). The values for $\alpha_{n}(\mathrm{Cu})$ can be well calculated and $C_{n}(B) = \alpha_{n}(\mathrm{Cu}) B^{2}$ with $\alpha_{n}(\mathrm{Cu}) = 14.7 \times 10^{-6}$\,JK/T$^{2}$mol (see SM)~\cite{hagino1994,kittaka2014}. After having subtracted $\alpha_{n}(\mathrm{Cu})$ we can plot the square root of the extracted $\alpha_{n}(\mathrm{Yb})$ coefficients together with $M(B)$. The values are indicated by black dots in the inset of Fig.~\ref{fig4} after having multiplied them by a constant factor 18.5. Using these points and the low-field magnetization at 45\,mK, it is possible to construct the full magnetization at 45\,mK (red dashed line). This curve allows us to estimate the $\alpha_{n}(\mathrm{Yb})$ coefficient for each field with a very good precision. We did then subtract the nuclear contribution $C_{n} = C_{n}(\mathrm{Yb}) + C_{n}(\mathrm{Cu})$ from all the data and the results are plotted in Fig.~\ref{fig4}: At $B \geq 5$\,T $C(T)/T$ is constant at low-$T$ indicating a Fermi liquid state.
However, for $B < 5$\,T an additional contribution to $C(T)/T$ appears and continuously increases with decreasing field. The fact that this contribution increases with decreasing temperature indicates that this has to include both electronic and nuclear degrees of freedom. We believe that this is a unique observation and shows that we do not have nuclear energy levels decoupled from the $4f$-electron levels. Instead there exists a coupling possibly because of the very low Kondo temperature~\cite{flouquet1978}. We are here in the 'slow-relaxation' regime and the electronuclear coupling is as large as several milliKelvin. The strong increase of $C(T)/T$ is then due to electronuclear quantum fluctuations at the QCP which results in a broadening of the electronuclear energy levels. This broadening is the effect of the slow relaxation and is seen in the temperature dependence of the specific heat, which instead of following a standard Schottky $C_{n}/T \propto 1/T^{3}$ function for well separated sharp energy levels, it follows a weaker power-law $C_{el-n} \propto 1/T^{2}$. 

It is possible to simulate this behavior with a simple resonant-level-model (RLM) presented in Refs.~\cite{schotte1975,desgranges1982} and often used to calculate the specific heat of Kondo systems in magnetic field. The model describes a two-energy-level system with level broadening $\Delta$ of Lorenz shape in magnetic field $B_{eff}$. This is illustrated in the inset of Fig.~\ref{fig3}. If $\Delta = 0$ or $B_{eff} \gg \Delta$ then $C_{n}/T \propto 1/T^{3}$ as in the Schottky case, but if $\Delta > 0$ and $\Delta \approx B_{eff}$ then $C_{n}/T \propto 1/T^{2}$ on the high-$T$ section, as we observe in \YCA. The strong $1/T^{2}$ increase of $C/T$ at low fields can be only explained by a nuclear contribution with a substantial broadening of the energy levels due to fluctuations of the $4f$-moments coupled to the nuclear moments. This is supported by the evolution to a $1/T^{3}$ behavior at large magnetic fields which suppress fluctuations and, therefore, it corresponds to $\Delta \rightarrow 0$. 

Even though it was not possible to carry out M\"ossbauer experiments on our sample to directly measure the relaxation rate $\tau_{J}$ of the Yb-nuclei, it is possible to estimate it from the $\mu$SR and NQR measurements done by Carretta \et~\cite{carretta2009}. To extract $1/T_{1}$($^{171,173}$Yb) we can scale equation 5 in Ref.~\cite{carretta2009} from $^{63}$Cu to $^{171}$Yb and $^{173}$Yb. The precise calculation is presented in the SM and we obtain for $^{173}$Yb:

\begin{equation}
\begin{aligned}
&\tau_{J}(0.1\,\mathrm{K})& = &~ 3.3 \times 10^{-10}\,\mathrm{s}&&\\
&\tau_{J}(1\,\mathrm{K})& = &~ 7.7 \times 10^{-10}\,\mathrm{s}.&&
\end{aligned}
\end{equation}
These values are close to the transition between the slow and fast relaxation regimes and therefore support a possible electronuclear coupling~\cite{nowik1968}. In fact, the hyperfine constant of the $^{170}$Yb isotope associated with a $\Gamma_{7}$ doublet is about 1\,GHz~\cite{bonville1984}. In M\"ossbauer experiments at 4.2\,K on the parent compound YbCu$_{4}$Au which is located away from the QCP with a $T_{\mathrm{N}} = 1$\,K (see inset of Fig.~\ref{fig1}), relaxation times of about 50\,GHz were measured~\cite{bonville1992}, i.e. only one order of magnitude larger than those expected at the QCP.

\begin{figure}[t]
	\begin{center}
		\includegraphics[width=\columnwidth]{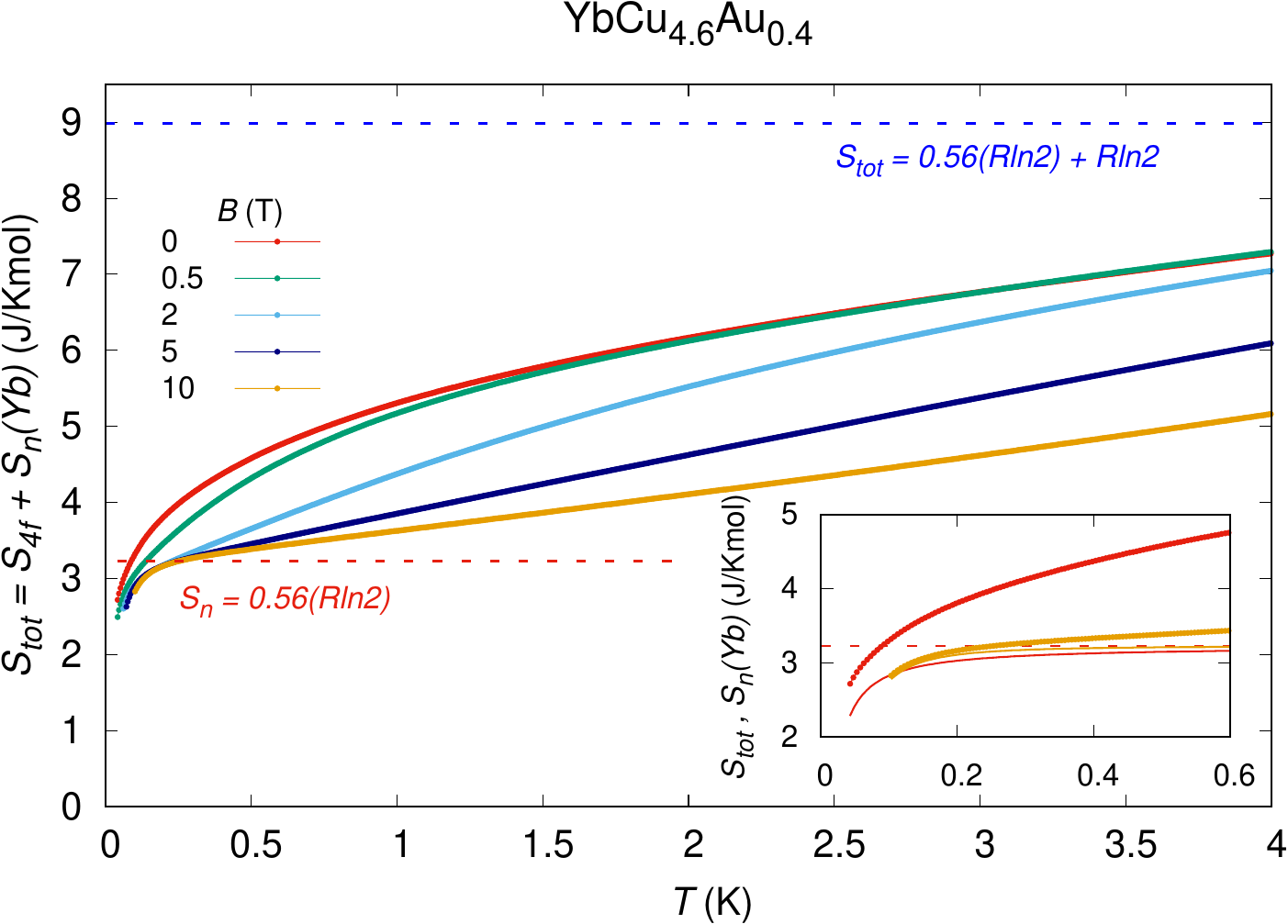}
		\caption{Total entropy $S_{tot}(T,B)$ curves for \YCA\ calculated as described in the text. They include the $4f$-electron entropy of the ground state $S_{4f} = R\ln2$ and the Yb nuclear entropy $S_{n}(\mathrm{Yb}) = 0.14 \cdot R\ln2+0.16 \cdot R\ln6 = 0.56 \cdot R\ln2$. The nuclear entropy of Cu has been subtracted. The inset shows $S_{tot}$ (points) and $S_{n}(\mathrm{Yb})$ (lines) at 0 and 10\,T.}
		\label{fig8}
	\end{center}
\end{figure}

From the experimental data and the low-$T$ fits in Fig.~\ref{fig3}, it is possible to estimate with good precision the evolution of the total entropy $S_{tot}(T,B) = S_{4f} + S_{n}(\mathrm{Yb})$ which include the $4f$-electron entropy of the ground state $S_{4f} = R\ln2$ and the Yb nuclear entropy $S_{n}(\mathrm{Yb}) = 0.14 \cdot R\ln2+0.16 \cdot R\ln6 = 0.56 \cdot R\ln2$. The nuclear entropy of Cu has been subtracted (cf. SM). $S_{tot}(T)$ and $S_{n}(T)$ are plotted in Fig.~\ref{fig8}. To estimate these curves we used the following procedure: We have evaluated the integral of the measured $C/T$ vs $T$ data starting from the lowest measured point $T_{low}$. For the data at high magnetic fields we have then subtracted the entropy of Cu atoms, which amounts to about 15\% at 10\,T, but is below 5\% at fields below 5\,T (see SM). Using the functions in Fig.~\ref{fig3} (dashed lines) we have evaluated the nuclear entropy for $T \geq T_{low}$. This yields, e.g., an entropy of 1\,J/Kmol for the $B = 0$ curve, which is about 1/3 of the total nuclear entropy of the Yb nuclei. The Yb nuclear entropy below $T_{low}$ is therefore the full entropy $0.56 \cdot R\ln2$ minus these values. We have also calculated the electronic entropy for $T \leq T_{low}$: To do this we used the constant values of $C/T$ after subtraction of the low-$T$ contributions according to the fit functions. 

We can now use the entropy plots and, for instance, estimate the Kondo temperature defined as $S(\frac{1}{2}T_{\textrm{K}}) = \frac{1}{2}R\ln 2$. Hence, we obtain $T_{\textrm{K}} \approx 4$\,K. The inset of the same figure, in which we plotted $S_{tot}$ and $S_{n}(\mathrm{Yb})$ for $B = 0$ and $B = 10$\,T, shows how the Yb nuclear entropy evolves in temperature reaching $0.56 \cdot R\ln2$  at about 0.6\,K for $B = 0$ and at 0.2\,K for $B = 10$\,T. This is compared with the total entropy. This gives us the energy scale of the electronuclear coupling, which is about 0.6\,K. It is only possible at high magnetic fields to fully disentangle the electronic and the nuclear contributions by suppressing the fluctuations.

In conclusion, the anomalous behavior of the $T$ and $B$-dependencies of the specific heat of the QC compound \YCA\ and the strong fluctuations observed in this material, strongly point to the presence of electronuclear coupling which results in a broadening of the electronuclear energy levels. This broadening is caused by the quantum critical fluctuations and is observed in the specific heat as $C/T \propto 1/T^{2}$ divergence. This is the first example of electronuclear quantum criticality in a metal.

One question may arise: Why is this effect observed in this compound and not in other Yb-based 'local-moment' systems with very low transition temperatures, like \YPS\ (\Tm\ = 250\,mK), \YPI\ (\Tm\ = 180\,mK)~\cite{gruner2014,jang2015} or \YCN\ (\Tm\ = 170\,mK)~\cite{sereni2018}? The basic difference is that systems like \YPS/In and \YCN\ are not at a QCP, actually rather away from it, and also do not show FM fluctuations. There are short-range correlations also in these materials, but their small ordering temperatures are due to a weak coupling of the order of 0.5\,K.

\begin{acknowledgments}
We are indebted to T. L\"uhmann for experimental support and P. Carretta for useful discussions. E. H. acknowledges support from Deutsche Forschungsgemeinschaft (DFG) for the CRC 1143 - project number 247310070, the CRC 88 - project E03 number 107745057 and for the W\"urzburg-Dresden cluster of excellence EXC 2147 ct.qmat Complexity and Topology in Quantum Matter - project number 390858490.
\end{acknowledgments}
\newpage
\noindent
\textbf{\Large Supplemental Material}
\renewcommand{\thefigure}{S\arabic{figure}}
\setcounter{figure}{0}
\section{Specific heat}
We show here all specific heat data measured on \YCA\ to emphasize the continuous evolution of the low-$T$ power-law behavior of $C/T$ from $C/T \propto T^{-0.75}$ at $B=0$ to $C/T \propto T^{-3}$
above 5\,T.
\begin{figure}[ht!]
	\begin{center}
		\includegraphics[width=\columnwidth]{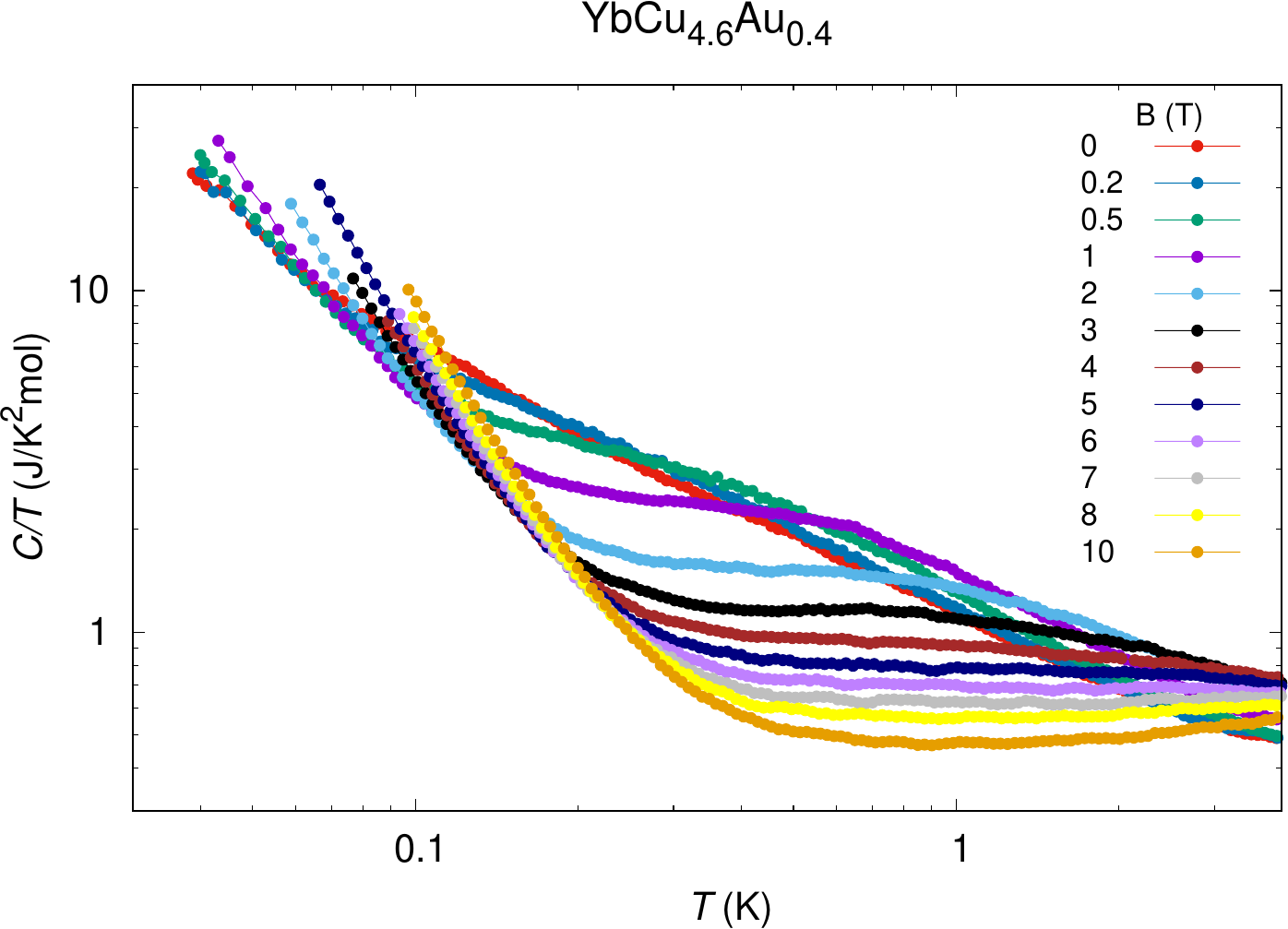}
		\caption{Full set of specific heat data taken on \YCA. $C/T$ follows a power-law function $T^{-0.75}$ at $B = 0$. With increasing $B$ the low-$T$ power-law function changes continuously to the standard $T^{-3}$ Schottky contribution expected for well separated nuclear energy levels.}
		\label{fig8_SI}
	\end{center}
\end{figure}
\section{SCR-theory and Specific heat capacity}
\begin{figure}[ht!]
	\begin{center}
		\includegraphics[width=\columnwidth]{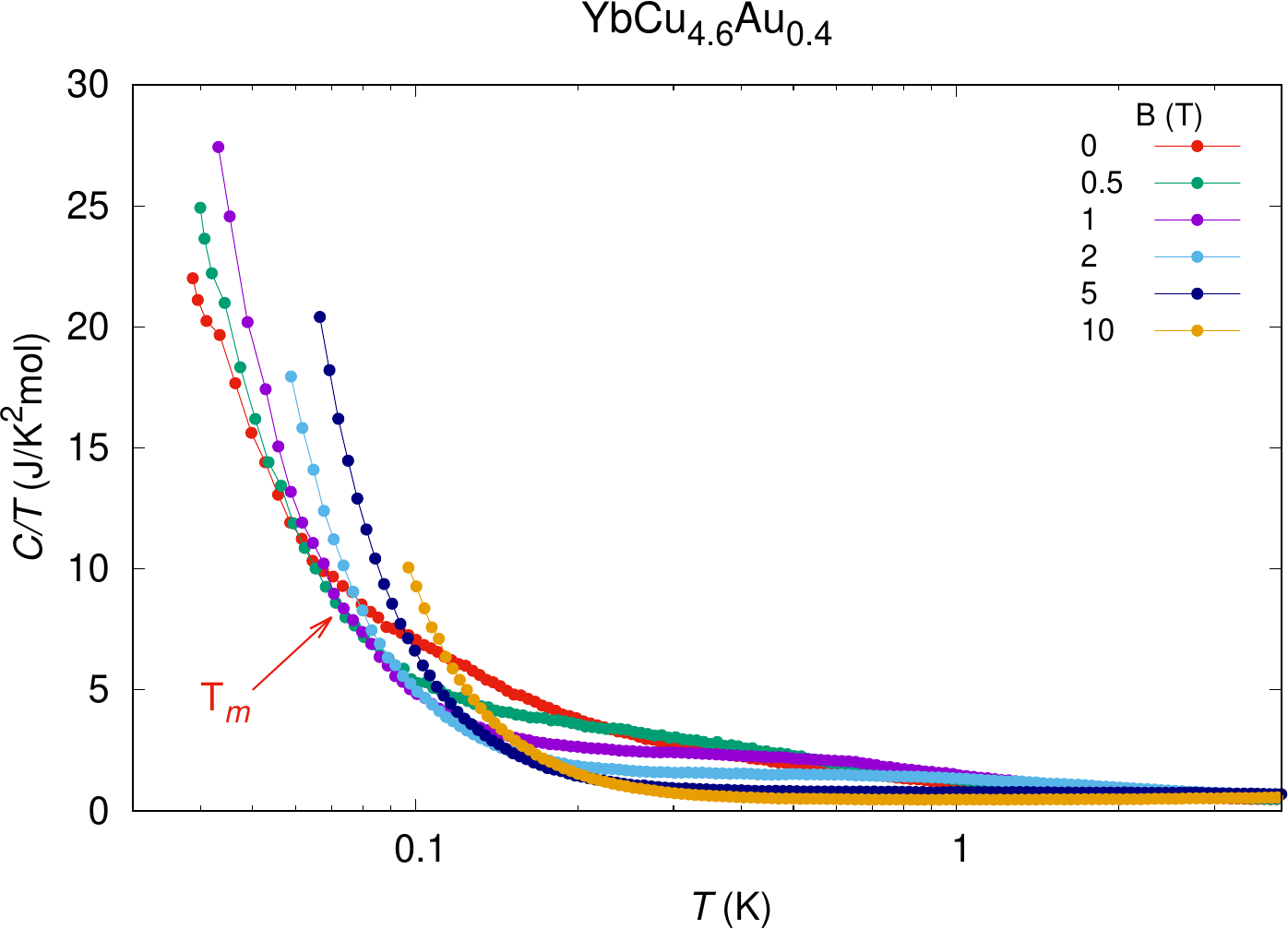}
		\caption{Specific heat capacity of \YCA\ plotted as $C(T)/T$ vs $\log(T)$ to emphasize the strong increase of $C(T)/T$ below 1\,K which follows a power-law function close to $T^{-0.75}$.}
		\label{fig9_SI}
	\end{center}
\end{figure}

$\mu$SR and NQR measurements on a sample of the series \YCAx\ with $x = 0.6$ have detected strong FM correlations at low temperature, the $T$-dependence of which could be well fitted with the SCR theory for a FM QCP~\cite{carretta2009}. In this theory the specific heat capacity is predicted to follow a logarithmic increase at temperatures lower than the characteristic energy scale of the spin fluctuations $T_{0}$~\cite{moriya1985}:
\begin{equation}\label{C}
\frac{C(T)}{T} = \delta \ln(T/T_{0}) + \beta T^{2}.
\end{equation} 
To demonstrate that this does not work for \YCA, we have plotted in Fig.~\ref{fig9_SI} $C(T)/T$ vs $\log(T)$ at zero and at selected magnetic fields (the same as those in Fig.~3 of the main text). $C(T)/T$ does not follow a logarithmic increase below 1\,K, but a power-law function close to $T^{-0.75}$. This is a much stronger enhancement than a $-\log(T)$.
\section{Calculation of the relaxation rate}
The characteristic energy of spin fluctuations $T_{0}$ can be directly measured by $\mu$SR and NQR. This was done by Carretta \et~\cite{carretta2009} on a sample of the same series \YCAx\ but with $x = 0.6$ (not exactly at the QCP) and by others in YbAuCu$_{4}$~\cite{koyama2002,yamamoto2008}.
$T_{0}$ is related to the spin-flip relaxation rate $\tau_{sf} = 1/T_{1}$ by the equation 5 in Ref.~\cite{carretta2009} which is derived from the SCR theory~\cite{moriya1985,maclaughlin1981}:
\begin{equation}\label{SCR}
\frac{1}{T_{1}} \simeq \gamma^{2}A^{2}\frac{3\hbar}{8\pi}\left(\frac{T}{T_{0}}\right)\frac{\chi_{s}}{N_{A}}.
\end{equation}
In this equation $1/T_{1}$, $\gamma$ and $A$ are the spin-lattice relaxation, the gyromagnetic ratio and the hyperfine coupling constant of $\mu^{+}$ or $^{63}$Cu in \YCA\ and $\chi_{s}$ is the magnetic susceptibility. 

We can calculate $T_{0}$ from $(1/T_{1})_{63}$ of the $^{63}$Cu NQR measurements at $T = 1$\,K bei Carretta \et~\cite{carretta2009} (similar values from Refs.~\cite{koyama2002,yamamoto2008}) and the following values:
\begin{itemize}
	\item $\gamma_{63} = 11.285$\,MHz/T
	\item $A_{63} = -0.63$\,kOe/\muB~\cite{koyama2002}
	\item $(1/T_{1})_{63}=(1/T_{1})_{\mu}/263=3/263$\,($\mu$s)$^{-1}$ =\\ 
	= 11407\,s$^{-1}$~\cite{carretta2009}
	\item $\chi_{s} = 5\times 10^{-6}$\,m$^{3}$/mol = (50/4$\pi$)\,J/T$^{2}$mol
	\item $h = 6.63\times 10^{-34}$\,Js
	\item \muB\ = $9.274 \times 10^{-24}$\,J/T
	\item $N_{A} = 6.023 \times 10^{23}$ mol$^{-1}$
\end{itemize}
We obtain the very low value: $T_{0} \approx 0.04\,\mathrm{mK}$.

Using Carretta \et\ data and this equation it is possible to calculate the relaxation rate $1/T_{1}$ expected for the Yb-nuclei rather precisely. As mentioned in the introduction of the main paper, this rate is the key to understand the low-$T$ behavior of \YCA. Relaxations rates are typically between $\tau_{J} \approx 10^{-8}$\,s (slow relaxation regime) to $\tau_{J} \approx 10^{-12}$\,s (fast relaxation regime)~\cite{nowik1968}. At slow rates, which are comparable to the nuclear Larmor period ($\tau_{n} \approx h/A$\kB), electrons can couple to the nuclei and this is what we believe it is happening in \YCA. At fast rates this coupling is prevented by the rapid fluctuations as it is observed in Kondo systems with high $T_{K}$.

To extract $1/T_{1}$($^{171,173}$Yb) we can scale Eq.~\ref{SCR} from $^{63}$Cu to $^{171}$Yb and $^{173}$Yb: The spin susceptibility $\chi_{s}$ is the susceptibility of the electronic $4f$-moments, i.e., it is the same for muon, Cu and Yb, and corresponds to the measured dc-susceptibility. In order to deduce $1/T_{1}$ of Yb from that of muon and $^{63}$Cu, one just needs the scale the square of the hyperfine couplings $A$ and the gyromagnetic ratios $\gamma$. We know from Ref.~\cite{carretta2009} that $(1/T_{1})_{\mu} = 263(1/T_{1})_{63}$, so we just need to scale values from $^{63}$Cu.

The hyperfine coupling relevant for Yb is well known and is the mean value between the on site hyperfine coupling given by experiment and theory, $A_{\mathrm{Yb}} = 110$\,T/\muB, which is inbetween the experimental $A_{\mathrm{Yb}} = 102$\,T/\muB~\cite{bonville1984,bonville1991,steppke2010} and the theoretical $A_{\mathrm{Yb}} = 120$\,T/\muB. The hyperfine coupling of Cu can be experimentally deduced from a plot of the Knight shift versus susceptibility. That has been done, e.g., by Koyama \et~\cite{koyama2002}. For $^{63}$Cu on the 16e site they deduced $A_{63} = - 0.063$\,T/\muB. So the ratio is $A_{\mathrm{Yb}}/A_{63}  = -1.746 \times 10^{3}$.

Gyromagnetic factors can be deduced from standard tables of NMR frequencies at a given field: At 2.3488\,T, $\nu_{63} = 26,505$\,MHz ($\nu_{65} = 28,394$\,MHz), $\nu_{\mathrm{^{171}Yb}} = 17.613$\,MHz and $\nu_{\mathrm{^{173}Yb}} = 4.852$\,MHz. We can calculate the rate for $^{173}$Yb because it is a spin $I = 5/2$, thus sligthly closer to $I = 3/2$ of $^{63}$Cu compared to the $I = 1/2$ of $^{171}$Yb. This yields a ratio of the gyromagnetic factors $\gamma_\mathrm{^{173}Yb}/\gamma_{63} = 0.183$.

Therefore:
\begin{equation}
\left(\frac{1}{T_{1}}\right)_{\mathrm{^{173}Yb}} = \left(\frac{1}{T_{1}}\right)_{\mu} \frac{(1.75 \times 10^3 \times 0.183)^{2}}{236}.
\end{equation}
If we take from Ref.~\cite{carretta2009} the values for $(1/T_{1})_{\mu} = 7 \times 10^{6}$\,s$^{-1}$ and $(1/T_{1})_{\mu} = 3 \times 10^{6}$\,s$^{-1}$ at 0.1\,K and 1\,K, respectively, we obtain:
\begin{equation}
\begin{aligned}
&\left(\frac{1}{T_{1}}\right)_{\mathrm{^{173}Yb}} & (0.1\,\mathrm{K}) &= 3 \times 10^{9}\,\mathrm{s}^{-1}&&\\
&\left(\frac{1}{T_{1}}\right)_{\mathrm{^{173}Yb}} & (1\,\mathrm{K}) &= 1.3 \times 10^{9}\,\mathrm{s}^{-1}.&&
\end{aligned}
\end{equation}

These values are close to the transition between the slow and fast relaxation regimes and therefore support a possible electronuclear coupling.  
\section{Nuclear contribution of copper}
The nuclear contribution of $^{63,65}$Cu ($I = 3/2$) with a total natural abundance of 100\% can be neglected at low fields but not at high fields. This topic is discussed, e.g., in Refs.~\cite{hagino1994,kittaka2014}.

The point group of the wyckoff position 16e for Cu in YbCu$_{4}$Au (space group F-43m) is trigonal whereas the one for Yb and Au is cubic. In \YCA, 0.6 of the total 4.6 Cu formula unit content goes into the Au cubic position~\cite{giovannini2005}. Hence, we do not expect in zero field a nuclear contribution to the specific heat from Yb, Au and 0.6 Cu quadrupole moments.

$^{63}$Cu ($I = 3/2$, 69.09\% n.a.) and $^{63}$Cu ($I = 3/2$, 30.91\% n.a.) have nuclear electric quadrupole moments $Q$ of about $-0.2 \times 10^{-24}$\,cm$^{2}$ and a splitting of the nuclear energy levels is allowed at $B = 0$ for 4 Cu atoms in the formula unit \YCA. In fact, the NQR frequency $\nu_{Q}(^{63}\mathrm{Cu}) = 8.96$\,MHz (0.43\,mK) was detected~\cite{koyama2002,carretta2009}. The value for $^{65}$Cu is very close. From this value we can calculate the $\alpha_{n}(B = 0)$ coefficient of the nuclear contribution to the specific heat, which is:
\begin{equation}
\alpha_{n}(B = 0) = \frac{R}{4}\nu_{Q}^{2} = 3.84 \times 10^{-7}\,\mathrm{JK/mol}.
\end{equation}
Considering the 4 Cu atoms in the formula unit with non-cubic point symmetry, we have $\alpha_{n}(B = 0, \mathrm{4\cdot Cu}) = 1.54 \times 10^{-6}$\,JK/mol which yields, for instance, a value for $C/T$ at 0.1\,K of:
\begin{equation}
C_{n}(B=0,0.1\,\mathrm{K})/T = 1.54 \times 10^{-3}\,\mathrm{J/K^{2}mol}
\end{equation}
which is definitely negligible (cf. Fig.~\ref{fig8_SI}).

In magnetic field, all 4.6 Cu atoms contribute to the Zeeman energy splitting of the nuclear levels. It is possible to calculate precisely the nuclear contribution to the specific heat caused by the Zeeman term in copper~\cite{hagino1994,kittaka2014}, but without knowing the field gradient at the Cu sites in \YCA\ the quadrupole contribution can be only guessed. We explain here how we have calculated these contributions.

Since the experimental data are dominated by the Yb contribution, we only consider theoretical calculations of the contribution of Cu-nuclear moments, as shown by Hagino \et~\cite{hagino1994}. In Refs.~\cite{hagino1994,kittaka2014} it is shown that theoretical values are very close to the experimental ones. The Zeeman term $C_{n}(B) = \alpha_{n}(\mathrm{Cu}) B^{2}$ can be taken from Ref.~\cite{kittaka2014} and is $\alpha_{n}(\mathrm{Cu}) = 6.4 \times 10^{-6}$\,JK/T$^{2}$mol for CeCu$_{2}$Si$_{2}$, so it is $\alpha_{n}(\mathrm{Cu}) = 14.7 \times 10^{-6}$\,JK/T$^{2}$mol for \YCA. At 10\,T it yields a contribution $C_{n}(\mathrm{Cu}) = 14.7 \times 10^{-4}$\,J/Kmol which is 16.7\% of the total measured $\alpha_{n} (\mathrm{Cu+Yb}) = 8.8 \times 10^{-3}$\,J/Kmol. At 5\,T this contribution is about 5\% and it is negligible at lower fields.
%
%
\section{Field dependence of ac-susceptibility}
\begin{figure}[t]
	\begin{center}
		\includegraphics[width=\columnwidth]{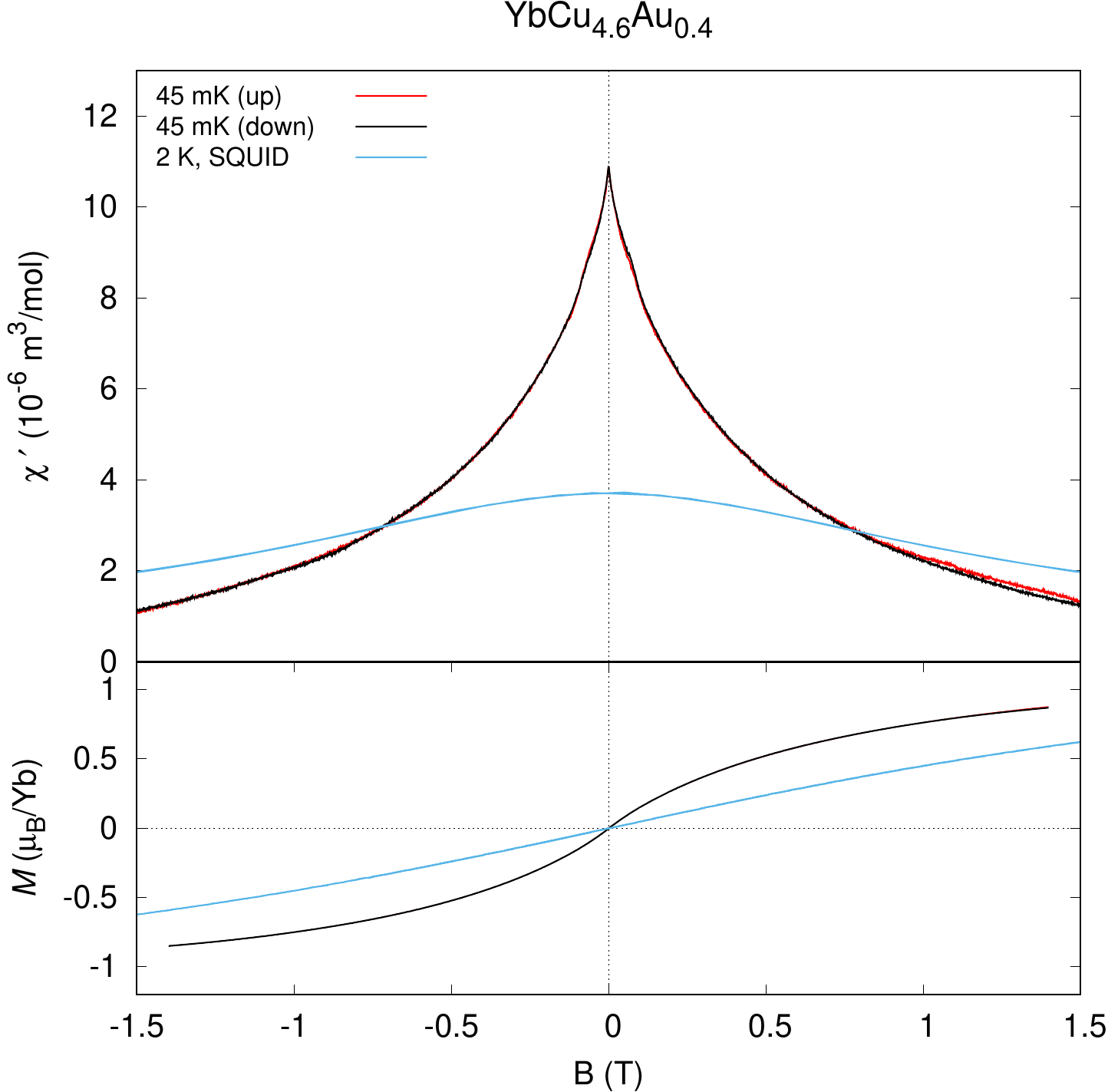}
		\caption{Magnetic field dependence of $\chi'(B)$ and magnetization $M(B)$ measured at 2\,K. The field sweeps were measured at 0.06\,T/min. The magnetization at 45\,mK was calculated by integrating $\chi'(B)$.}
		\label{fig11_SI}
	\end{center}
\end{figure}
We have also measured $\chi'(B)$ at 45\,mK between -1.5 and 1.5\,T to look for possible features or hysteresis. This is shown in Fig~\ref{fig11_SI}. There are neither clear features nor hysteresis in both sweep-up and sweep-down curves confirming the presence of a very weak short-range ordering. This is also shown in the integral, i.e. the magnetization plotted in the lower panel of the same figure.

So, the broad transition with a tiny amount of entropy quenched and the small critical field $< 15$\,mT point to weak short-range ordering, the long-range order is prevented by the strong fluctuations. Very similar behavior was observed in a close related system \YCN~\cite{sereni2018}.
\bibliography{banda_prx_2023.bib}
\bibliographystyle{apsrev4-2}
\end{document}